\documentclass[aps,pre,twocolumn,showkeys,floatfix,groupeaddress,longbibliography]{revtex4-2}

\usepackage{amsmath,amsfonts,amssymb}
\usepackage{graphicx,psfrag}
\usepackage[caption=false]{subfig}
\usepackage{verbatim}
\usepackage{xcolor}

\graphicspath{{Plots/}}
\begin{document}

%\preprint{}

\title{Over-Relaxation in Diffusive Integer Lattice Gas}


\author{Kyle Strand}
\email[]{kyle.t.strand@ndsu.edu}
\author{Alexander J. Wagner}
\email[]{alexander.wagner@ndsu.edu}
\homepage[]{www.ndsu.edu/pubweb/~carswagn}
%\thanks{}
\affiliation{Department of Physics, North Dakota State University, Fargo, ND 58108}

\date{\today}

\begin{abstract}
One of the most striking draw-backs of standard lattice gas methods over lattice Boltzmann methods is a much more limited range of transport parameters that can be achieved. It is common for lattice Boltzmann methods to use over-relaxation to achieve arbitrarily small transport parameters in the hydrodynamic equations. Here, we show that it is possible to implement over-relaxation for integer lattice gases. For simplicity we focus here on lattice gases for the diffusion equation. We demonstrate that adding a flipping operation to lattice gases results in a multi-relaxation time lattice Boltzmann scheme with over-relaxation in the Boltzmann limit.
\end{abstract}


\maketitle

\section{Introduction}
Lattice Boltzmann methods have emerged as a highly successful numerical method for many areas of fluid flow and beyond. However, particularly for fluctuating systems, the discrete nature of earlier lattice gas methods seems much more appropriate, since fluctuations in nature are directly related to the discreteness of matter. Original lattice gas approaches were inferior to lattice Boltzmann methods in several respects.
However, recent developments in integer lattice gases by Blommel \textit{et al.} \cite{blommel2018integerreal} showed that many of the artifacts of traditional Boolean lattice gas methods \cite{frisch1986,doolen1991} could be overcome by allowing for integer occupation numbers. One of the remaining shortcomings of the integer lattice gas methods is that the resulting transport coefficients have a more limited range than the transport coefficients that can be a achieved using lattice Boltzmann methods. Lattice Boltzmann methods routinely use a collision operator that over-relaxes the local distributions which allows them to achieve arbitrarily small transport coefficients. This motivated us to further investigate the possibility of achieving over-relaxation in integer lattice gas methods.

Integer lattice gas (ILG), as developed by Blommel \textit{et al.} \cite{blommel2018integerreal} provided a template for extending traditional lattice gas cellular automaton methods \cite{frisch1986,doolen1991}. Traditionally, lattice gases only allowed a single particle per lattice node, however Chopard \textit{et al.} presented a scheme for multi-particle ILG \cite{chopard_droz_1998}. In recent years, we have seen a reemergence of research in practical applications of lattice gas methods \cite{parsa2017latticereal,parsa2018validitypub}.  Blommel presented a novel way of allowing any integer number of particles to occupy any node and be consistently re-distributed through binary collisions that conserved both mass and momentum. This implementation mitigated issues such as Galilean invariance, which plagued lattice gas methods, so that they would become equivalent to the corresponding lattice Boltzmann methods. However, it was computationally expensive thus limiting the practical application of such a model. Seekins \textit{et al.} \cite{seekins2022integerreal}, inspired by Boghosian and Chopard \textit{et al.} \cite{boghosian1997, chopard_droz_1998, chopardprl1998}, presented a modification to the collision operator which, instead of following the defined collision rules step-by-step, sampled a probability distribution to arrive at the same results as presented in Blommel's original model. 

The success of the Seekins' sampling collision operator in terms of computational practicality sheds new light on what could be achieved using ILG methods. One key draw-back of lattice gas methods is that they do not appear to allow for over-relaxation of the collision operator. Such over-relaxation is routinely used in lattice Boltzmann methods to achieve lower transport coefficients. This is particularly helpful for hydrodynamic simulations at high Reynolds numbers, which are helped significantly if low viscosities are possible. The idea of over-relaxation is that instead of local collisions moving a local distribution closer to local equilibrium the effect is to over-shoot this approach and land beyond the equilibrium distribution.  Such over-relaxation has been shown by B\"osch and Karlin to be disconnected from the standard kinetic theory domain from which lattice Boltzmann methods are typically derived \cite{bosch2013}. More recently Pachalieva was able to show that over-relaxation can also be obtained by a simple coarse-graining of Molecular Dynamics simulations \cite{pachalieva2021connectingreal}, giving a more direct link between over-relaxation and a physical system. This inspired us to question the generally held idea that over-relaxation is not possible in lattice gas methods, and in particular are revisiting this question for integer lattice gas methods. 

In this manuscript, we present a simple and effective method which successfully performs over-relaxation in a diffusive integer lattice gas. We extend the sampling collision operator presented by Seekins to incorporate over-relaxation through a simple permutation of particles during the collision. In section \ref{sec:ilg}, we introduce the basics of the integer lattice gas method. Section \ref{sec:overrelax} provides the necessary extensions for implementing over-relaxation in ILG methods. We derive the Boltzmann average for the system in section \ref{sec:boltzmannapprox} showing its correspondence with lattice Boltzmann methods and we show that we can derive the diffusion equation from the equation of motion for the over-relaxed integer lattice gas. In section \ref{sec:results}, we verify the validity of our refined collision operator incorporating over-relaxation.

%Just a brief overview of integer lattice gas
 \section{Integer Lattice Gas}\label{sec:ilg}
 We review here briefly the integer lattice Boltzmann method of Seekins \textit{et al.} \cite{seekins2022integerreal}. It consists of an underlying regular lattice where neighboring lattice points are connected through lattice velocities $\{v_i \Delta t\}$. With each of these lattice velocities at each lattice point we associate integer occupation numbers $n_i(x,t)$ that evolve through the lattice gas evolution equation
\begin{equation}
     n_i(x + v_i \Delta t, t + \Delta t) = n_i(x,t) + \Xi_i(\{n_i\}).
     \label{eq:ilgeq}
\end{equation}
Here $\Xi_i$ is a collision operator that redistributes the particles at each lattice site. This collision operator is stochastic by nature and must obey all local conservation laws. In our case only mass is conserved, so we require 
\begin{equation}
     \sum_i \Xi_i = 0.
\end{equation}
Seekins \textit{et al.} \cite{seekins2022integerreal} introduced a collision operator that picked a fraction $\omega$ of particles at random and redistributed them to occupation number $n_i$ with a probability $w_i$, where the $w_i$ are the familiar weight functions used in the definition of lattice Boltzmann equilibrium distributions \cite{Qian_1993}. The details of the algorithm that allows this to be done efficiently will not be discussed here, but are detailed in the publication cited above.
 
For this lattice gas we can derive a lattice Boltzmann average through
\begin{equation}
    f_i(x,t) = \langle n_i(x,t) \rangle,
\end{equation}
 where the average $\langle \cdots \rangle$ implies a non-equilibrium average over all possible realizations of the stochastic lattice gas. The same average is applied to the lattice gas collision operator to obtain the lattice Boltzmann collision operator
 \begin{equation}
     \Omega_i = \langle \Xi_i \rangle.
 \end{equation}
 This lattice Boltzmann collision operator was shown to be of the form
 \begin{equation}
     \Omega_i = \omega (f_i^0-f_i)
     \label{eqn:Omega}
 \end{equation}
 where the local equilibrium distribution is given by
 \begin{equation}
     f_i^0(x,t) = \rho(x,t) w_i
 \end{equation}
 with the local density
 \begin{equation}
     \rho(x,t) = \sum_i f_i(x,t).
 \end{equation}
 The resulting lattice Boltzmann equation
 \begin{equation}
     f_i(x+v_i\Delta t,t+\Delta t) = f_i(x,t)+\omega[f_i^0(x,t)-f_i(x,t)],
 \end{equation}
 can then be shown to have the diffusion equation as its hydrodynamic limit:
 \begin{equation}
     \partial_t \rho(x,t) = \nabla D \nabla \rho(x,t),
 \end{equation}
 where the diffusion constant is given by
\begin{equation}
    D=\left(\frac{1}{\omega}-\frac{1}{2}\right)\theta,
\end{equation}
with 
\begin{equation}
    \theta=\sum_i w_i v_i^2.
    \label{eqn:Thetadef}
\end{equation}
In lattice Boltzmann simulations that are used as numerical methods in their own right, values of $\omega\in\{0,2\}$ are routinely used, but in lattice gas implementations the definition of $\omega$ as a probability limits its range to $\omega\in\{0,1\}$. This limits the usefulness of lattice gas methods compared to their lattice Boltzmann counterparts.

\section{Over-relaxation in a lattice gas}
\label{sec:overrelax}
The contribution of this paper is to show that it is indeed possible to construct lattice gas methods that can access the $\omega \in \{1,2\}$ range often used in lattice Boltzmann approaches. This is of interest not only for the diffusive systems considered here, but also for hydrodynamic systems, which are the main forte of lattice Boltzmann methods. In those systems the transport coefficient of interest is the viscosity, and obtaining low values for the viscosity is essential for simulations of systems with high Reynolds numbers. In this manuscript we present a proof of principle that we hope to extend to those even more important hydrodynamic LB models in the near future.

The range of $\omega\in\{1,2\}$ is referred to as over-relaxation because the lattice Boltzmann collision operator will overshoot the local equilibrium distribution in the relaxation process. $\omega=1$ corresponds to full relaxation, where the distributions reach local equilibrium in each step, and $\omega<1$ implies under-relaxation. That there is a difficulty of deriving over-relaxation from, say, a continuous Boltzmann equation was shown by B\"osch \textit{et al.} \cite{bosch2013}. But aside from its obvious practical utility it has now been shown by Pachalieva \textit{et al.} \cite{pachalieva2021connectingreal} that over-relaxation in lattice Boltzmann can also be obtained by coarse-graining Molecular Dynamics simulations. 

As indicated by B\"osch \textit{et al.} over-relaxation cannot be obtained through a continuous extension of the collision process. We propose here to augment the collision process with a flipping operation such that 
\begin{equation}
    F_i(n_i) = n_{-i}.
    \label{eq:flip}
\end{equation}
where we interpret negative indices such that $v_{-i}\rightarrow -v_i$. This leads to the lattice gas evolution equation
\begin{equation}
    n_i(x+v_i \Delta t, t+\Delta t) = F_i(n_i)+\Xi_i(\{F_i(n_i)\}).
    \label{eqn:LGflip}
\end{equation}

Heuristically such a flipping operation will send particles back along the direction they just came from, and it is reasonable to expect that this operation will completely suppress diffusion. It is therefore reasonable to expect that this operation on its own will lead to something resembling $\omega=2$, \textit{i.e.} full over-relaxation. A closer examination of the collision operator Eqn. \ref{eqn:Omega} shows that this is not the full story as we will examine below.

This flipping operation is then augmented with an additional collision with a collision fraction $\omega^* \in \{0,1\}$. In the limiting case of $\omega^*=0$, we only apply the flipping operation, leading to the case of full over-relaxation, and the limiting case of $\omega^*=1$ means that all particles are re-distributed, making the flipping operation moot and leading to full relaxation. In the next section we will derive the Boltzmann limit of this augmented lattice gas and show that it indeed corresponds to a lattice Boltzmann method with over-relaxation.

\section{Boltzmann approximation}\label{sec:boltzmannapprox}
We will now derive the Boltzmann average of the lattice gas including the flip operator of Eqn. (\ref{eqn:LGflip}). This is most easily accomplished by separating the LB operation into a collision and a streaming step and then transform the collision term into a moment space where the flip-operation has a very simple interpretation. We obtain
\begin{equation}
    f_i(x+v_i\Delta t,t+\Delta t) = F_i[f_i(x,t)]+\omega\{f_i^0(x,t)-F_i[f_i(x,t)]\}.
\end{equation}
Following the procedure presented by Wagner \textit{et al.} \cite{wagner2016fluctuatingreal}, we transform the distribution functions into moment space by defining a transformation matrix $m_i^a$ with which we obtain occupation numbers in moment space
\begin{align}
    M^a &= \sum_i m_i^a f_i.
\end{align}
The transformation matrix is orthogonal with respect to the Hermite norm defined through
\begin{align}
    \sum_i m_i^a w_i m_i^b = \delta^{ab},\\
    \sum_a m_i^a w_j m_j^a = \delta_{ij}.
\end{align}
This allows us to obtain the $f_i$ from the $M^a$ through
\begin{align}
    f_i &= \sum_a w_i m_i^a M_a.
\end{align}
When designing a transformation matrix it is customary that the first moments should correspond to the conserved quantities, the following moments to the hydrodynamic quantities, and the remainder will represent so-called ghost modes, i.e. quantities that do not enter the hydrodynamic limit.

For a simple one dimensional model with three velocities $\{v_i\}=\{-1,0,1\}$ (D1Q3) this transformation matrix is written as \cite{wagner2016fluctuatingreal}
\begin{equation}
    m_i^a = \left(
    \begin{array}{ccc}
    1 & 1 & 1 \\
    -\sqrt{\frac{1}{\theta}} &0 & \sqrt{\frac{1}{\theta}} \\
     \sqrt{\frac{1-\theta}{\theta}}&-\sqrt{\frac{\theta}{1-\theta}} & \sqrt{\frac{1-\theta}{\theta}}
    \end{array}
    \right),
    \label{eq:transform}
\end{equation}
where $\theta$ was defined in Eqn. (\ref{eqn:Thetadef}).
It is useful to give the moments $M^a$ separate names related to their physical significance:
\begin{equation}
    M^a = \left(
    \begin{array}{c}
    \rho \\
    j \\
    \Pi
    \end{array}
    \label{eq:momentspacevector}
    \right),
\end{equation}
where $\rho$ is the particle density, $j$ is the current density, and $\Pi$ is related to the energy density moments.
A particularly nice property of this transformation matrix is that (in general) the value of the non-conserved quantities of the equilibrium distribution in moment space are zero \cite{kaehler2013derivationreal}:
\begin{equation}
    M^{a,0} = \sum_i m_i^a f_i^0 = 
    \left(
    \begin{array}{c}
    \rho \\
    0\\
    0
    \end{array}
    \label{eq:momentspacef0}
    \right),
\end{equation}
We can now separate out the effect of the flipping operator and the collision process. The flipping operator has a very simple representation in moment space:
\begin{equation}
    F(M^a) = \left(
    \begin{array}{c}
    \rho \\
    -j \\
    \Pi
    \end{array}
    \label{eq:momentspaceFm}
    \right),
\end{equation}
and in general all even velocity moments are unaffected by the flipping operation and all odd velocity moments acquire a negative sign. The effect of the collision is likewise simple: conserved quantities are unaffected and non-conserved quantities are multiplied by the fraction $\omega$. So we can write the effect of the collision operator in moment space as
\begin{equation}
    M^{a}=M^a+\Omega^{a}=\left(
    \begin{array}{c}
    0 \\
    -(1-\omega) j \\
    (1-\omega)\Pi
    \end{array}
    \label{eq:momentspaceOm}
    \right).
\end{equation}
This is equivalent to defining a new 
\begin{equation}
    \omega^j=2-\omega,
    \label{eqn:omegaj}
\end{equation} 
so that we get the more usual
\begin{equation}
    M^{a,*}=\left(
    \begin{array}{c}
    0 \\
    (1-\omega^j) j \\
    (1-\omega)\Pi
    \end{array}
    \label{eq:momentspaceOm}
    \right).
\end{equation}
With this we can write the lattice Boltzmann equation corresponding to the integer lattice gas with the flipping operation as
\begin{equation}
    f_i(x+v_i\Delta t,t+\Delta t) = \sum_a w_i m_i^a (1-\omega^a) m_j^a(f_j^0-f^j),
    \label{eqn:MRTLB}
\end{equation}
where we introduced the $\omega^a$ notation to refer to $(\omega^\rho,\omega^j,\omega)$ and $\omega^\rho$ is arbitrary.
The key result is that this has the form of a standard MRT lattice Boltzmann equation. In the case where $\omega^j \in [1,2]$, over-relaxation is observed, which achieves the primary goal of the flipping operation, $F(M^a)$.

According to the derivation of the hydrodynamic limit of the lattice Boltzmann equation in Eqn. (\ref{eqn:MRTLB}) (see \textit{e.g.} Kaehler \textit{et al.} \cite{kaehler2013derivationreal}), we obtain the diffusion equation
\begin{equation}
    \partial_t \rho = -D \nabla^2\rho,
    \label{eq:diffusioneqn}
\end{equation}
in which we defined a diffusion constant
\begin{equation}
    D=\theta\left(\frac{1}{\omega^j} - \frac{1}{2}\right).
\end{equation}
As shown by Sorenson \textit{et al.} \cite{sorenson2022}, the evolution of the densities follows the diffusion equations for features with wavelength $\lambda \gtrapprox 10 \pi/\omega$. In the following we choose $\lambda=320$ which is well in this regime.

\section{Results}\label{sec:results}
In order to verify that this diffusive implementation does replicate Eqn. (\ref{eq:diffusioneqn}) in the hydrodynamic limit, we analyze a system with a known analytic solution \cite{blommel2018integerreal,seekins2022integerreal}. We impose the system with a sine wave as the density profile which takes the form 
\begin{equation}
    \rho(x,0) = N^{ave}\left[1+\sin\left(\frac{2\pi x}{L}\right)\right].
    \label{eq:sineinit}
\end{equation}
Here, $N^{ave}$ is the average number of particles which exist at each node on the lattice and $L$ is the size of the one-dimensional lattice. The time-evolution of this system has the analytical solution
\begin{align}
    \rho(x,t) &= N^{ave}\left[1+\sin\left(\frac{2\pi x}{L_x}\right)\exp{\left(-\frac{4\pi^2 D t}{L^2}\right)}\right]\\
              &= N^{ave} + A^{th}(t)\sin\left(\frac{2\pi x}{L}\right),
\end{align}
where we have a definition for the decay of the amplitude 
\begin{equation}
    A^{th}(t) = N^{ave}\exp{\left(-\frac{4\pi^2 Dt}{L^2}\right)}.
    \label{eq:ampdecay}
\end{equation}

\begin{figure}
    \includegraphics[width=\columnwidth,clip=true]{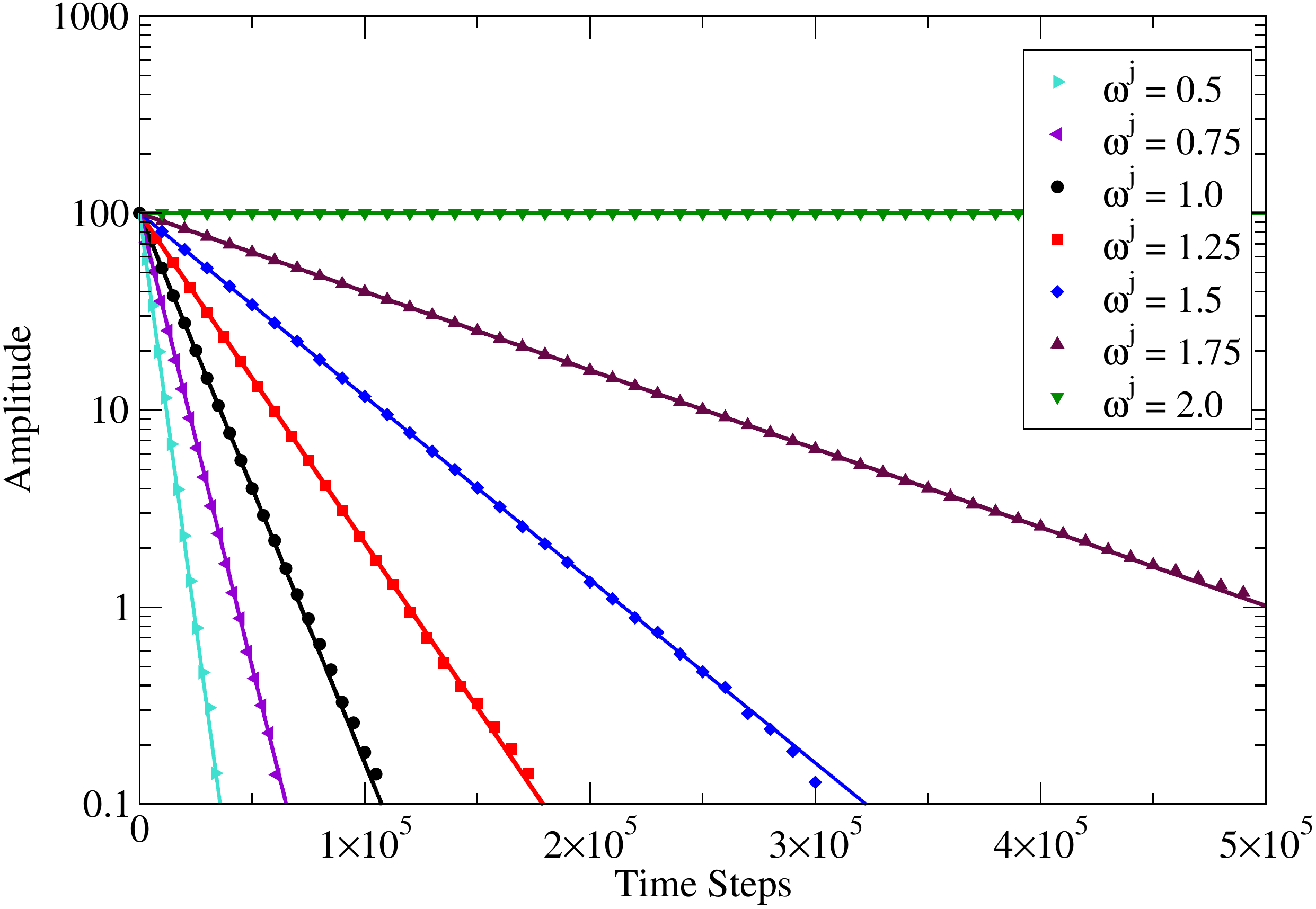}
    \caption{Decay of the amplitude of an initial sine wave with varying $\omega^j$ values. $\omega^j > 1$ is in the over-relaxed regime. We see good agreement between the measured simulation data (symbols) and the theoretical prediction (solid lines) from Eqn. (\ref{eq:ampdecay}) both inside and outside the over-relaxation regime. This data was a result of the average of 500 individual simulations on a D1Q3 lattice with size $L=320$ and $N^{ave} = 100$.}
    \label{AmpDecay}
\end{figure}

An issue arises due to the fact that Eqns. (\ref{eq:sineinit}-\ref{eq:ampdecay}) are continuous, the ILG methods are discrete by nature, and $n_i \in \mathbb{Z}$. To remedy this, we can impose our initial density profile which is from a sinusoidal probability distribution
\begin{equation}
    P(\rho) = N^{ave}\left[1+\sin\left(\frac{2\pi x}{L}\right)\right].
\end{equation}
This can be performed by choosing Poisson distributed random numbers for the occupation numbers with an expectation value $w_i P(\rho)$ based on the weighting of the system \cite{seekins2022integerreal}. This method allows us to model continuous functions as fully integer valued which properly aligns with the discrete nature of the ILG methods. In the same manner as Blommel, we are able to acquire the amplitude of the profile at any point by
\begin{equation}
    A^{LG}(t) = \frac{\sum_x \sin\left(\frac{2\pi x}{L}\right) N(x,t)}{\sum_x \sin^2\left(\frac{2\pi x}{L}\right)}.
\end{equation}
This measured amplitude can then be compared to the theoretical prediction in Eqn. (\ref{eq:ampdecay}). We illustrate this comparison in Figure \ref{AmpDecay} where we see excellent agreement between the measured amplitudes and the theoretical prediction for the decay of the sinusoidal profile. For the values $1 < \omega^j < 2$ in the over-relaxation regime, we find very good agreement between the measured simulation and the theoretical prediction. The values without the flipping operation had previously been verified by Seekins for $\omega \leq 1$ and are shown for completeness. It is interesting to point out the behavior which occurs when $\omega^j = 2$. In this case, our collision probability is $\omega = 0$. Here, the flipping operation is guaranteed to permute all particles back to their original position through the collision with a probability of 1. This in combination with the streaming step will cause all the particles in the system to continually permute which will not evolve the system at all. This data was acquired using a D1Q3 lattice with a size of $L = 320$ with $N^{ave}=100$ particles per lattice node averaged over 500 individual simulations.

%\section{Investigations of Correspondence with Lattice Boltzmann Methods}

%{\color{red}{We will discuss how over-relaxed LB corresponds with the LG stuff previously.}}

\section{Conclusions}
We have presented a new method for successfully performing over-relaxation in diffusive integer lattice gas models, which had previously been thought to be impossible. This method introduces a simple permutation of the occupation numbers within the system to over shoot local equilibrium. This is made possible by defining an effective collision probability which nullifies the mathematically impossibility of utilizing probabilities greater than 1. This works in tandem with the sampling collision operator presented by Seekins, but it is also generally possible to implement on any collision operator. The ability to utilize over-relaxation in integer lattices gases will increase the usefulness and practicality of the method. The example of diffusion has provided a pathway in which we intend to develop a fully realized over-relaxed hydrodynamic integer lattice gas.
\bibliography{AW,LG}

\end{document}